\documentclass[english,aps,prc,superscriptaddress,amsmath,amssymb,showpacs]{revtex4}
\usepackage[T1]{fontenc}
\usepackage[latin9]{inputenc}
\usepackage{amstext}

\makeatletter

\providecommand{\tabularnewline}{\\}

\@ifundefined{textcolor}{}
{%
 \definecolor{BLACK}{gray}{0}
 \definecolor{WHITE}{gray}{1}
 \definecolor{RED}{rgb}{1,0,0}
 \definecolor{GREEN}{rgb}{0,1,0}
 \definecolor{BLUE}{rgb}{0,0,1}
 \definecolor{CYAN}{cmyk}{1,0,0,0}
 \definecolor{MAGENTA}{cmyk}{0,1,0,0}
 \definecolor{YELLOW}{cmyk}{0,0,1,0}
 }

\makeatother

\usepackage{babel}
\begin{document}

\title{Isobaric analog state in the f$_{7/2}$ and g$_{9/2}$ shells}

\author{L. Zamick}

\author{A. Escuderos}

\affiliation{Department of Physics and Astronomy, Rutgers University, Piscataway,
New Jersey, 08854}

\author{S.J.Q. Robinson}

\affiliation{Department of Physics, Millsaps College, Jackson, Mississippi, 39210}

\author{Y. Y. Sharon}

\affiliation{Department of Physics and Astronomy, Rutgers University, Piscataway,
New Jersey, 08854}

\author{M.W. Kirson}

\affiliation{Department of Particle Physics and Astrophysics,Weizmann Institute
of Science, 76100 Rehovot, Israel}
\begin{abstract}
Calculations are performed for energies of isobaric analog states
with isospins T=2 and T=3 in regions where they have been found experimentally
e.g. f-p shell, and regions where they have not yet been found e.g.
g$_{9/2}$ near Z=50,N=50. We consider two approaches--one using binding
energy formulas and Coulomb energies contained therein and the other
using shell model calculations. It is noted that some (but not all)
calculations yield very low excitation energies for the J=0$^{+}$T=2
isobaric analog state in $^{96}$Ag. 
\end{abstract}
\maketitle
If there were no violation of charge independence, the binding energy
of the $^{96}$Pd ground state ($J=0^{+}$, $T=2$) would be identical
to the binding energy of the analog state, also $J=0^{+}$, $T=2$,
in $^{96}$Ag. But, since that is not the case in real life, the excitation
energy of the $J=0^{+}$, $T=2$ state in $^{96}$Ag is given by 
\begin{equation}
E^{*}(J=0^{+},T=2)=BE(^{96}\text{Ag})-BE(^{96}\text{Pd})+V_{C}\,,\label{eq:exc}
\end{equation}
 where the $BE$s are the binding energies and $V_{C}$ includes all
charge-independence violating effects. The binding energies can be
obtained from the latest mass evaluation \cite{am11} and we assume
that $V_{C}$ arises from the Coulomb interaction, which must be estimated.

We use the classical form of the Coulomb energy 
\begin{equation}
E_{C}=\alpha_{C}Z^{2}/A^{1/3}\,,\label{eq:ecd}
\end{equation}
 supplemented by an exchange Coulomb term 
\begin{equation}
E_{xC}=\alpha_{xC}Z^{4/3}/A^{1/3}\,,\label{eq:ecx}
\end{equation}
 where $\alpha_{C}$ and $\alpha_{xC}$ are coefficients to be obtained
from appropriate data. Several sources were compared. The simplest
is the Bethe-Weizsäcker semi-empirical mass formula \cite{key-3,mwk},
which produces $\alpha_{C}=0.691$ MeV, $\alpha_{xC}=0$ from a fit
of a four-term semi-empirical mass formula to the measured masses.
An extended, ten-term mass formula \cite{mwk} produces $\alpha_{C}=0.774$
MeV and $\alpha_{xC}=-2.22$ MeV from a similar fit. The best mass
formulation currently available is the Duflo-Zuker approach \cite{key-4,mwdz}
with up to 33 parameters fitted to the mass data. It includes a unified
Coulomb term 
\begin{equation}
E_{C}^{DZ}=\alpha_{C}\frac{Z(Z-1)-0.76[Z(Z-1)]^{2/3}}{A^{1/3}\left[1-\frac{(N-Z)^{2}}{4A^{2}}\right]}\label{eq:dzc}
\end{equation}
 and the best fits to the data have $\alpha_{C}=0.700$ MeV.

Binding energy differences of mirror nuclei, together with Coulomb
displacement energies, can be fitted to differences of $E_{C}$ and
$E_{xC}$ (eqs.(\ref{eq:ecd}),(\ref{eq:ecx})), from which $\alpha_{C}=0.717$
MeV and $\alpha_{xC}=-0.928$ MeV \cite{mwk}. The formula of Anderson
et al.~\cite{awm65}: 
\begin{equation}
V_{C}=E_{1}\overline{Z}/A^{1/3}+E_{2}\,,\label{eq:vc}
\end{equation}
 where $\overline{Z}=(Z_{1}+Z_{2})/2$, is a semi-empirical representation
of the same data, as far as it was known at the time. Anderson et
al.~\cite{awm65} list several sets of values of $E_{1}$ and $E_{2}$.
We here use the average values $E_{1}=1.441$~MeV and $E_{2}=-1.06$~MeV.

Table~\ref{tab:coule} compares the Coulomb energy estimates, using
the different prescriptions presented above, for a number of nuclei
of interest for this discussion of analog state excitation energy.
Though estimates of the total Coulomb energy can vary strongly between
prescriptions, the differences which are relevant to the analog states
show much less variability. In particular, estimates which are based
on fits to mirror nuclei and Coulomb displacement energies agree very
closely among themselves. The Anderson et al fit has stood the test
of time remarkably well.

\begin{table}
\caption{\label{tab:coule}Coulomb energy estimates for some nuclei, in MeV.
Lines labeled $A=$ give the differences of the two preceding lines.}

\begin{ruledtabular} %
\begin{tabular}{cccccc}
Nucleus  & Bethe-Weizsäcker  & Ten-term  & Duflo-Zuker  & Mirror/CDE  & Anderson et al \tabularnewline
\hline 
$^{44}$Sc  & 86.318  & 60.253  & 74.872  & 74.336  & \tabularnewline
$^{44}$Ca  & 78.293  & 53.558  & 67.586  & 66.968  & \tabularnewline
$A=44$  & 8.025  & 6.694  & 7.285  & 7.368  & 7.308 \tabularnewline
\hline 
$^{46}$Sc  & 85.048  & 59.366  & 73.872  & 73.242  & \tabularnewline
$^{46}$Ca  & 77.141  & 52.771  & 66.738  & 65.983  & \tabularnewline
$A=46$  & 7.907  & 6.596  & 7.134  & 7.259  & 7.185 \tabularnewline
\hline 
$^{52}$Mn  & 115.706  & 86.126  & 102.432  & 101.885  & \tabularnewline
$^{52}$Cr  & 106.635  & 78.268  & 94.079  & 93.435  & \tabularnewline
$A=52$  & 9.071  & 7.858  & 8.353  & 8.450  & 8.399 \tabularnewline
\hline 
$^{60}$Cu  & 148.442  & 115.748  & 133.491  & 132.907  & \tabularnewline
$^{60}$Ni  & 138.381  & 106.788  & 124.048  & 123.433  & \tabularnewline
$A=60$  & 10.061  & 8.960  & 9.362 & 9.474  & 9.430 \tabularnewline
\hline 
$^{94}$Rh  & 307.747  & 266.563  & 286.532  & 286.658  & \tabularnewline
$^{94}$Ru  & 294.221  & 253.719  & 273.683  & 273.588  & \tabularnewline
$A=94$  & 13.526  & 12.843  & 12.849  & 13.070  & 13.043 \tabularnewline
\hline 
$^{96}$Ag  & 333.362  & 291.169  & 311.153  & 311.530  & \tabularnewline
$^{96}$Pd  & 319.328  & 277.773  & 297.738  & 297.939  & \tabularnewline
$A=96$  & 14.035  & 13.396  & 13.415  & 13.591  & 13.574 \tabularnewline
\end{tabular}\end{ruledtabular} 
\end{table}

With relatively stable Coulomb energy differences in hand and with
experimental binding energies, we are able to compute, using eq.(\ref{eq:exc}),
predicted excitation energies of analog states, and can compare the
results with measured excitation energies, where they exist. We show
in Table~\ref{tab:exc} results for various nuclei, some for which
the excitation energy of the analog state is known and some for which
it is not. The binding energy differences \mbox{BE(Z,N)-BE(Z+1,N-1)}
are taken from Ref.~\cite{am11}, the Coulomb energy differences
from the Anderson et al semi-empirical fit \cite{awm65}.

\begin{table}
\caption{\label{tab:exc}Excitation energies of isobaric analog states in MeV.}

\begin{ruledtabular} %
\begin{tabular}{ccccccc}
NUCLEUS  & Binding Energy Difference  & Coulomb Energy  & Excitation Energy  & Single $j$  & Large space  & Experiment \tabularnewline
\hline 
$^{44}$Sc  & 4.435  & 7.308  & 2.873  & 3.047\footnotemark[1]  & 3.418\footnotemark[2]  & 2.779 \tabularnewline
$^{46}$Sc  & 2.160  & 7.185  & 5.024  & 4.949\footnotemark[1]  & 5.250\footnotemark[2]  & 5.022 \tabularnewline
$^{52}$Mn  & 5.494  & 8.399  & 2.905  & 2.774  & 2.7307  & 2.926 \tabularnewline
$^{60}$Cu  & 6.910  & 9.430  & 2.520  & 2.235  & 2.726\footnotemark[2]\footnotemark[7]  & 2.536 \tabularnewline
$^{94}$Rh  & 10.458  & 13.043  & 2.585  & 1.990\footnotemark[3]  & 3.266\footnotemark[4]  & \tabularnewline
 &  &  &  & 2.048\footnotemark[3]  & 2.879\footnotemark[6]  & \tabularnewline
$^{96}$Ag  & 12.453  & 13.574  & 1.121  & 0.900\footnotemark[3]  & 1.9167\footnotemark[4]  & \tabularnewline
 &  &  &  & 0.842\footnotemark[5]  & 1.640\footnotemark[6]  & \tabularnewline
\end{tabular}\end{ruledtabular} \footnotetext[1]{Escuderos, Zamick, Bayman (2005)~\cite{ezb05}.}
\footnotetext[2]{GXPF1 interaction~\cite{hobm04}.} \footnotetext[3]{Zamick
and Escuderos (2012)~\cite{ze12}.} \footnotetext[4]{jj44b interaction~\cite{bl-un}.}
\footnotetext[5]{CCGI interaction~\cite{ze12,ccgi12}.} \footnotetext[6]{JUN45
interaction~\cite{jun45} } \footnotetext[7]{truncated model space,
allowing 4 nucleons excited from the f$_{7/2}$ subshell} 
\end{table}

In all four cases where the excitation energy of the analog state
is known, our prediction agrees with the experimental value within
100 keV, and for three of them, within 25 keV. The fact that the analog
state and Coulomb arguments work well in known cases gives us confidence
that we can use these for the unknown case of $^{96}$Ag, where we
predict an excitation energy just slightly above 1 MeV. Turning things
around, if the isobaric analog state were found, then we might have
a better constraint on what the binding energy is.

We can compare our predicted excitation energies with selected calculations
in the literature (included in Table \ref{tab:exc}). We look at shell-model
calculations of two basic kinds --- single-$j$ or large--space ---
and with various effective interactions. For $^{44}$Sc and $^{46}$Sc,
single-$j$-shell results ($f_{7/2}$)~\cite{ezb05} are respectively
3.047 and 4.949~MeV, as compared with Table~\ref{tab:exc}'s excitation
energies of 2.873 and 5.024~MeV. The large space results are also
shown. In$^{52}$Mn there is reasonable agreement between predicted,
single-$j$, large space and experiment. In the small space for $^{60}$Cu
(p$_{3/2}$) we can use a particle-hole transformation to get the
spectrum of this nucleus from the spectrum of $^{58}$Cu since three
p$_{3/2}$ neutrons can be regarded as a single neutron hole. This
gives a value of 2.235 MeV as compared with the experimental value
of 2.536 MeV.

For $^{96}$Ag single-$j$-shell results~\cite{ze12} are 0.900~MeV
with INTd and 0.842~MeV with the CCGI interaction~\cite{ze12,ccgi12}.
These are lower than the excitation energy in Table~\ref{tab:exc}
of 1.121~MeV. There are also large scale calculations with the jj44b~\cite{bl-un}
interaction for $^{96}$Ag---the result is 1.917~MeV, significantly
larger than the predicted value. In $^{94}$Rh the jj44b interaction
yields 3.266~MeV, larger than Table~\ref{tab:exc}'s predicted value
of 2.585~MeV. The large space calculations with JUN45 are qualitatively
similar.The single-$j$ INTd and CCGI results are lower, 1.990 MeV
and 2.048 MeV respectively.

Although it is clearly preferable to base predictions of the excitation
energy of the analog state on experimentally measured binding energies,
it may become necessary to use binding energies derived from mass
formulas where data is unavailable. To this end, we check how susceptible
these predictions are to various mass formulas. We tested the mass
formulas used above to obtain Coulomb energy differences --- a 5--term
Bethe-Weizsäcker formula (the standard four terms, supplemented with
a pairing term), its 10--term extension, and the 33-parameter Duflo--Zuker
mass formulation. The results for the analog states are presented
in Table \ref{tab:msfrm}. In all cases, the Coulomb energy differences
were obtained from the respective binding energy formulas.

\begin{table}
\caption{\label{tab:msfrm}Excitation energies of isobaric analog states (in
MeV) based on mass formulas, for comparison with predicted analog
state energy in Table \ref{tab:exc}.}

\begin{ruledtabular} %
\begin{tabular}{ccccccc}
NUCLEUS  & 5-term  & 5-term  & 10-term  & 10-term  & Duflo-Zuker  & Duflo-Zuker \tabularnewline
 & Binding Energy  & Analog Energy  & Binding Energy  & Analog Energy  & Binding Energy  & Analog Energy \tabularnewline
\hline 
$^{44}$Sc  & 5.499  & 2.526  & 4.768  & 1.926  & 4.934  & 2.351\tabularnewline
$^{46}$Sc  & 1.650  & 6.257  & 2.035  & 4.624  & 2.440  & 4.694\tabularnewline
$^{52}$Mn  & 7.200  & 1.871  & 6.460  & 1.398  & 6.472  & 1.881\tabularnewline
$^{60}$Cu  & 8.653  & 1.408  & 7.968  & 0.992  & 7.015 & 2.347\tabularnewline
$^{94}$Rh  & 11.208  & 2.318  & 11.361  & 1.482  & 10.837  & 2.012\tabularnewline
$^{96}$Ag  & 13.668  & 0.367  & 13.453  & -0.057  & 12.885  & 0.530\tabularnewline
\end{tabular}\end{ruledtabular} 
\end{table}

The Duflo-Zuker results are closet to the predictions based on the
atomic mass evaluation (fourth column of Table \ref{tab:exc}),.As
might have been expected, mass formulas with smaller rms deviations
from the measured data are better predictors of the analog state excitation
energy. Even so the calculated value for $^{52}$Mn (1.881 MeV) is
considerably lower than the experimental value (2.926 MeV).

Why study isobaric analog states? One reason has to do with the strange
dualism in nuclear structure that has emerged over the years. For
the most part calculations of the excited states of nuclei have been
performed with little mind to the binding energies or saturation properties.
On the other hand binding energies and nuclear densities are addressed
in Hartree-Fock calculations with interactions for which it makes
no sense to calculate nuclear spectra. With isobaric analog state
energies we have an in your face confrontation of these two approaches.
As shown in Eq.~(\ref{eq:exc}) one needs good binding energies and
good Coulomb energies to correctly predict the excitation energies
of these states. As has been noted in ref \cite{ze12} with the shell
model one can get very impressive fits for many energy levels in say
$^{96}$Ag but no one has even tried to calculate the energy of the
isobaric analog state here until now. Hopefully our work will stimulate
trying to get a unified approach in which both the spectra and bulk
properties of nuclei are treated in a unified manner.

Another point of interest is the possibility that in some region of
the periodic table the T=2 isobaric analog state would become the
ground state. In Table \ref{tab:msfrm} the 10 point formula yields
such a result. If this were indeed the case there would be a drastic
difference in the decay mode of the nucleus in question. Instead of
the usual decay mode -- electron capture -- one would now have an
allowed Fermi transition. This would lead to a much shorter lifetime
and could influence how the elements evolve. This does not occur in
$^{96}$Ag where it is known that the decay mode is electron capture
but it might occur in heavier nuclei. For example Z=66, A=132 closes
the h$_{11/2}$ shell. Consider a very proton rich nucleus with A=128
and Z=63, a nucleus with 3proton holes and one neutron hole. The excttation
energies for 5- term, 10-term and Duflo-Zuker are respectively -0.009,
1.251 and 0.366 MeV.

Before leaving we would briefly like to defend out inclusion of small
space calculations in Table \ref{tab:exc}. There is a precedent for
this in the work of Talmi and collaborators \cite{key-1}. They obtain
excellent agreement with binding energies in several regions, e.g.
the Ca isotopes, using a single j-shell formula but with phenomenological
parameters.. We here adopt the same philosophy of obtaining two body
matrix elements from experiment. Matrix elements form experiment implicitly
contain not all but many nuclear correlations. 

In view of the differing results of shell model calculations and mass
formulas it would be of great interest to measure the excitation energies
of isobaric analog states in the $g_{9/2}$ region. We hope that this
work will encourage experimentalists to look not only for the surprisingly
neglected $J=0^{+}$ isobaric analog states in $^{94}$Rh and $^{96}$Ag,
but also for other such states throughout this region. 
\begin{acknowledgments}
We thank Klaus Blaum for his help. \end{acknowledgments}

\end{document}